\documentstyle[12pt]{article}

\begin{document}

\baselineskip 0.9cm

\vskip 2.0cm
\begin{center}
{\Large\bf           Three \vspace{2mm}Generations or More\\
                     for an Attractive Gravity ?}

\vskip 1.5cm

\large{Masahiro Tanaka}
\vskip 0.75cm
Department of Physics, Tohoku University, Sendai, 980  Japan
\vskip 1cm
\end{center}
\vskip 1.2cm

\begin{abstract}

  We calculate the induced Einstein action from supersymmetric models in
general space-time. Supersymmetric models consist of two kinds of
supermultiplets, called
scalar supermultiplets and vector supermultiplets, respectively. We show that
the vector
multiplets generate a negative Newton constant, while the scalar multiplets
 a positive one. Then we find that the positivity of Newton constant depends
on the ratio of the number of scalar multiplets to one of vector multiplets.
If we apply this result
to two hopeful supersymmetric unified
 models: one is minimal
SUSY standard model and another is minimal SUSY SU(5) GUT, we are led to the
conclusion that we need more than or equal to three generations of quark and
lepton multiplets to have a positive Newton constant, i.e., an attractive
gravity.

\end{abstract}

\newpage


\section{Introduction}

A long-standing unsolved problem in the field of high energy physics is
how one can decscribe the quantum theory of gravity satisfactorily. In such
a situation,
about a quarter century ago some people began to consider that a gravity
is only induced through the vacuum fluctuations of matter and gauge
fields $\cite{sa,a&c&m&t,a1}$, i.e., not
an elementary field. People call such a theory of gravity
 induced gravity or, more tasteful name, pregeometry.
In this model both the Newton constant and the cosmological constant are
determined by a
physical cut-off of some underlying theory. The quartically
 divergent cosmological constant, however, is opposite to our obsevations.
With this
fact in mind we introduce supersymmetry (SUSY) to the underlying theory,
because it is well known that SUSY gives rise to cancellations between
curvature independent vacuum fluctuations of fermions and those of bosons
$\cite{z,a}$, so leads to vanishing cosmological constant. Besides, SUSY, more
percisely supergravity, decides
the coefficient $\xi $ of conformally coupled term $-\xi {\cal R}AA^*$,
which the induced Einstein action drastically depends on via Seeley-DeWitt
coefficients, of some scalar field $A$. As a result there is a predictability
in our result. Also from the viewpoint of particle physics, SUSY makes
unified theories very hopeful as 'beyond the standard model'.

In what follows we consider that the candidates of an underlying theory of
induced gravity are supersymmetric models with other three forces, strong
 and electroweak interactions. Then we calculate the vacuum fluctuations of
both scalar multiplets and vector ones which are ingredinets of
supersymmetric unified models, and see what their induced Einstein actions
are. Finally, we apply these results to the two supersymmetric unified models
and conclude that we need three or more generations to obtain an attractive
induced gravity.

This paper is organized as follows. In chapter 2 we review the DeWitt-
Schwinger technique for the evaluation of the effective action. In chapter 3
and 4 we evaluate the induced Einstein action from the scalar multiplets and
the vector ones, respectively.
In chapter 5 we study the induced Einstein action from the underlying
theory with both scalar and vector multiplets. Then we apply this to two
unified models and obtain the suggestion that we need three or more
generations to have an attractive gravity.


\section{Effective Action and DeWitt-Schwinger Technique }

\indent
First we review how to evaluate effective actions, vacuum
fluctuations in our case, using DeWitt-Schwinger technique $\cite{d&s,dw}$.
Note that we calculate the effective action in Euclidean curved space-time
in order to
regularize this action by proper time cut-off $\cite{a1,d&s,sc}$.
A merit of proper time cut-off regularization is that it does not break
 SUSY so long as we are concerning about
divergent part, as we shall see below. And in the regularization we can
naturally identify the quartically divergent and curvature independent
term with cosmological constant. Likewise, we identify quadratically
divergent term linear in curvature with Einstein action.

We begin with the generating functional
$Z[g_{mn}]$:
\begin{eqnarray}
Z[\bar{\phi},g_{mn}]=\int {\cal D}\phi e^{-S[\phi ;\bar{\phi},g_{mn}]} ,
\label{2,1}
\end{eqnarray}
where $g_{mn}$ is the background gravitational field\footnote{We mean
auxiliary field in the underlying theory by background field.},
 and $\phi $ and
$\bar{\phi}$
represent all the quantum fields and classical ones but the background
gravitational field, respectively. Then the
effective action $W[\bar{\phi},g_{mn}]$
is defined as follows:
\begin{eqnarray}
Z[\bar{\phi},g_{mn}]=e^{-W[\bar{\phi},g_{mn}]}.
\label{2,2}
\end{eqnarray}
At one-loop level the effective action is formally written as follows:
\begin{eqnarray}
W[\bar{\phi},g_{mn}]=S[\bar{\phi},g_{mn}]\pm \frac{1}{2}\ln {\rm Det}
\Delta [\bar{\phi},g_{mn}] \; ,
\label{2,3}
\end{eqnarray}
where
\begin{eqnarray}
\Delta [\bar{\phi},g_{mn}]=-\Box +m^2[\bar{\phi},g_{mn}]-\xi {\cal R}.
\label{2,4}
\end{eqnarray}
$m$ and $\xi $ are a mass parameter and a conformal coupling to the background
gravitational field, respectively, and ${\cal R}$ is the
scalar curvature of background gravitational field. The signs above correspond
to fermion and boson, respectively. Using the identity
\begin{eqnarray}
\frac{1}{\Delta}=\int _0^{\infty}ds \; e^{-s\Delta}
\label{2,5}
\end{eqnarray}
we obtain
\begin{eqnarray}
\pm \frac{1}{2}{\rm Tr}\, {\rm Ln}\, \Delta = \pm \frac{1}{2}\int _0^{\infty}
\frac{ds}{s}{\rm Tr}\, e^{-s\Delta}\pm \mbox{ field independent constant}.
\label{2,6}
\end{eqnarray}
We set this field independent constant zero.\footnote{ In SUSY case
these constants cancel one another.}
The trace part of the integrand in eq.$(\ref{2,6})$ is
expressed in terms of heat kernel as follows
\begin{eqnarray}
{\rm Tr}\, e^{-s\Delta}=\int d^4x\sqrt{g}K(x,x;s) ,
\label{2,7}
\end{eqnarray}
where the heat kernel $K(x,x';s)=<x\mid e^{-s\Delta}\mid x'>$
and satifies the equations
\begin{eqnarray}
& &\frac{\partial}{\partial s}K(x,x';s)+\Delta K(x,x';s)=0\nonumber\\
& &K(x,x';0)=\delta (x,x').
\label{2,8}
\end{eqnarray}
General solution of the equations $(\ref{2,8})$ can be written
in terms of asymptotic series
\begin{eqnarray}
K(x,x';s)=\frac{1}{(4\pi s)^2}e^{-m^2s-\sigma (x,x')/2s}\sqrt{D(x,x')}
\sum _{n=0}^{\infty}E_ns^{n/2},
\label{2,9}
\end{eqnarray}
where $\sigma (x,x')$ is the half of square of geodesic length between $x$ and
$x'$, and $D(x,x')$ is the covariant Van Vleck-Morette determinant defined by
\begin{eqnarray}
D(x,x')=\frac{1}{\sqrt{g}}\det \{ -\partial _m \partial '_n\sigma (x,x')\}
\sqrt{g}.
\label{2,10}
\end{eqnarray}
In the coincidence limit $\sigma (x,x')\to 0$ and $D(x,x')\to 1$, one can
find one-loop part of the effective action, the second term of r.h.s. of
eq. $(\ref{2,3})$:
\begin{eqnarray}
W_{1-loop}=\pm \int d^4x\sqrt{g}\frac{1}{2(4\pi)^2}\sum _{n=0}^{\infty}
\int _0^{\infty}ds\; e^{-m^2s}
s^{n/2-3}E_n.
\label{2,11}
\end{eqnarray}
We separate the one-loop part of the effective action into the divergent part
${\cal L}_D$
and the finite part ${\cal L}_F$ :
\begin{eqnarray}
W_{1-loop}=\pm \int d^4x\frac{1}{2(4\pi)^2}({\cal L}_D+{\cal L}_F).
\label{2,12}
\end{eqnarray}
In order to regularize ${\cal L}_D$ and give it physical meanings, let us
introduce an
ultraviolet cut-off $\Lambda$, then we obtain
\begin{eqnarray}
& &g^{-1/2}{\cal L}_D=\sum _{n=0}^4E_n\int _{1/\Lambda ^2}^{\infty}ds\;
e^{-m^2s}s^{n/2-3}
\label{2,13}\\
& &g^{-1/2}{\cal L}_F=\sum _{n=5}^{\infty}\Gamma \Bigl( \frac{n}{2}-2\Bigr)
E_n(m^2)^{2-n/2}.
\label{2,14}
\end{eqnarray}
We can easily evaluate the divergent part, especially in the limit
$\Lambda \to \infty $
\begin{eqnarray}
g^{-1/2}{\cal L}_D&=&\frac{1}{2}E_0\Bigl[ \Lambda ^4-2m^2\Lambda ^2-m^4\Bigl(
\gamma
-\frac{3}{2}-\ln \frac{\Lambda ^2}{m^2}\Bigr)\Bigr] \nonumber\\
& &+\frac{2}{3}E_1(\Lambda ^3-3\Lambda m^2+2\sqrt{\pi}m^3)+
E_2\Bigl[ \Lambda ^2+m^2(\gamma -1)-m^2\ln \frac{\Lambda ^2}{m^2}\Bigr]
\nonumber\\
& &+2E_3(\Lambda -\sqrt{\pi}m)+E_4\Bigl( \ln \frac{\Lambda ^2}{m^2}-\gamma
\Bigr) .
\label{2,15}
\end{eqnarray}
For a manifold without boundary $E_{2n+1}=0$ and $E_{2n}=a_n$ where $a_n$
are well-known Seeley-DeWitt coefficients$\cite{d&s}$.
Non-vanishing first two coefficients are
\begin{eqnarray}
& &E_0=a_0=1\\
& &E_2=a_1=(\xi -1/6){\cal R}
\label{2,16}
\end{eqnarray}
a degree of freedom.
Here we find that in a manifold without boundary the coefficients of
logarithmically divergent terms are same as that of divergent terms in
dimensional regularization, and curvature independent terms of quartic
divergence and quadratic one vanish through Bose-Fermi cancellations.
After all it gives supersymmetric results as far as we are concerning on
cosmological constant and Einstein action.


\section{Induced Einstein Action from Scalar Multiplets}

In order to see the induced Einstein action purely from scalar multiplets,
we deal with
a curved space-time version of Wess-Zumino model $\cite{w&b}$ which consists of
scalar multiplets and an {\em auxiliary} gravity multiplet. We want to treat
unified models far below so called Planck scale ($M_P=1/\sqrt{8\pi G}$) and,
as we can see
from the eqs.$(\ref{2,15})$ and $(\ref{2,16})$, to obtain the
cosmological constant and Einstein action as fourth and second power of
cut-off, Planck scale, respectively. So we can ommit the
terms of $O(1/M_P)$ from the starting Lagrangian. With this fact in mind we
start from
the renormalizable part of an N=1\footnote{
N=1 means that there is only one Majorana spinor
field as gravitino.}  supergravity (locally supersymmetric )
Lagrangian $\cite{w&b}$ {\it without} terms as follows
\footnote{ This separation is justified
 when we treat the theory far below Planck scale and superpotenital without
linear term.} :
\begin{eqnarray}
 & &\frac{1}{8\pi G}\Biggl[ -\frac{1}{2}{\cal R}
-\frac{1}{3}MM^*+\frac{1}{3}b^ab_a \Biggr]
+\frac{i}{2}\epsilon ^{klmn}\bar{\psi}_k\gamma _5
\gamma _l\tilde{\cal D}_m\psi _n
\nonumber\\
& &-c\Biggl[ \frac{1}{8\pi G}M-\bar{\psi}_{La}\sigma ^{ab}\psi _{Lb}\Biggr]
-c^*\Biggl[ \frac{1}{8\pi G}M^*-\bar{\psi}_{Ra}\sigma ^{ab}\psi _{Rb}\Biggr] ,
\label{3,0}
\end{eqnarray}
where $c$ is a complex constant, $M(M^*)$ and $b_a$ are auxiliary fileds and
$\psi _n$ is the gravitino which is the gauge field of SUSY\footnote{
Accurately we mean no more than dimension-four parts of the supergravity
by SUSY,
supersymmetry, in this paper.} We may neglect $b_a\sim O(1/M_P^2)$ in the
above.
Above terms
$(\ref{3,0})$ include the
kinetic terms of graviton and gravitino, and cosmological terms of
supergravity, which are supersymmetric extension of Einstein action with
cosmological constant. Starting from the supergravity Lagrangian without
above terms
$(\ref{3,0})$, we obtain as the quantum corrections the induced action of
kinetic terms of graviton
and gravitino, and
(probably vanishing) cosmological constant. In other words, through the
quantum fluctuations of elementary fields we obtain the same
kinds of terms as those $(\ref{3,0})$, which we discarded at the underlying
Lagrangian level. Note that $G$ is would-be Newton constant.

We are here interested only in bosonic part of the induced Einstein
supergravity action, so we can set auxiliary gravitino field zero. In other
words, we can set the background space-time bosonic.
In calculating the induced Einstein action, we can confine ourselves in free
theory because we are only interested in vacuum fluctuations at one-loop
order. In practice we invoked
supergravity only to decide conformal
coupling $-\frac{1}{6}{\cal R}AA^*$. The decision of conformal coupling of
scalar field is equivalent to that of Einstein action  because the conformal
coupling of other fields (spinor and vector fields) to gravity is known
that $\xi =1/4$.
In Euclidean bosonic curved space-time, Lagrangian written in terms of
 a scalar
multiplet with supersymmetric mass $m$ are $\cite{w&b}$
\begin{eqnarray}
g^{-1/2}{\cal L}&=&
 \partial _mA^*\cdot \partial ^mA
+\frac{1}{2}\bar{\Psi}\gamma ^m{\cal D}_m\Psi -FF^*
\nonumber\\
& &+\frac{1}{3}(MA^*F+M^*AF^*)-\Bigl( \frac{1}{6}{\cal R}+\frac{1}{9}MM^*
\Bigr) AA^*
\nonumber\\
& &-m\Bigl \{ (AF+A^*F^*)-\frac{1}{2}\bar{\Psi}\Psi
-\frac{1}{2}(M^*A^2+MA^{*2})\Bigr \}
\label{3,1}
\end{eqnarray}
where $A$ is a complex scalar, $\Psi $ is a
Majorana spinor, and $F$ is a complex auxiliary scalar field.
And ${\cal D}_m$ is
covariant derivatives, ${\cal R}$ is the
scalar curvature and $M$ is auxiliary field of gravitational field.
Here we take the term with $mM$, $mM^*$ and $M^*M$ for mass terms like one
with $m^2$.
{}From the Lagrangian above we find the first two Seeley-DeWitt
coefficients which are
\begin{eqnarray}
& &a_0=2,
\label{3,2}\\
& &a_1=0,
\label{3,3}
\end{eqnarray}
for a complex scalar, and
\begin{eqnarray}
& &a_0=2
\label{3,4}\\
& &a_1=\frac{1}{6}{\cal R}
\label{3,5}
\end{eqnarray}
 for a Majorana spinor.

Then, for a Wess-Zumino model, we can easily
evaluate the vacuum fluctuation using the above coefficients.
The induced Einstein
action in Euclidean curved space-time is
\begin{eqnarray}
W_S=\int d^4x\sqrt{g}\frac{1}{2(4\pi)^2}\Bigl[ \frac{1}{6}{\cal R}\Lambda ^2
\Bigr] .
\label{3,6}
\end{eqnarray}
The sign of curvature term is same as one of $\cite{a}$. We find that,
without gauge
fields we always obtain a positive Newton constant.


\section{Induced Einstein Action from Vector (Gauge) Multiplets}

We will next study the induced action from an SU($N$) or a U($1$)
gauge multiplet in general
space-time, because it is also
contained in realistic unified models. As we mentioned in the previous
chapter,
we are interested only in vacuum energy density at one-loop order, so only
non-interacting parts are relevant here.
The pure supersymmetric Yang-Mills (renormalizable) Lagrangian in Wess-Zumino
gauge
$\cite{w&b}$ is
\begin{eqnarray}
g^{-1/2}{\cal L}=\frac{1}{4}F^{(a)}_{mn}F^{(a)mn}
+\frac{1}{2}\bar{\lambda }^{(a)}\gamma ^m\tilde{\cal D}_m\lambda ^{(a)}
-\frac{1}{2}D^{(a)2}
\label{4,1}
\end{eqnarray}
where $D^{(a)}=0$ from the equation of motion and $\tilde{\cal D}_m$ is
general coordinate
and gauge covariant derivatives. The non-interacting parts of Lagrangian
$(\ref{4,1})$ is
\begin{eqnarray}
g^{-1/2}{\cal L}_{non-int}
=\frac{1}{4}\Bigl[ \partial _mv^{(a)}_n-\partial _nv^{(a)}_m\Bigr]
\Bigl[ \partial ^mv^{(a)n}-\partial ^nv^{(a)m}\Bigr]
+\frac{1}{2}\bar{\lambda }^{(a)}\gamma ^m{\cal D}_m\lambda ^{(a)} .
\label{4,2}
\end{eqnarray}

Then, the non-interacting Lagrangian $(\ref{4,2})$ is the Lagrangian of
$ (N^2-1)$ independent supersymmetric electromagnetisms. Next step is to
quantize this theory. In order to do that we add the gauge fixing term and
corresponding ghost term to Lagrangian $(\ref{4,2})$ as everyone always does.

\begin{eqnarray}
g^{-1/2}{\cal L}_{tot}&=&
\frac{1}{4 }\Bigl[ \partial _mv^{(a)}_n-\partial _nv^{(a)}_m\Bigr]
\Bigl[ \partial ^mv^{(a)n}-\partial ^nv^{(a)m}\Bigr]
+\frac{1}{2}\bar{\lambda }^{(a)}\gamma ^m{\cal D}_m\lambda ^{(a)}\nonumber\\
& &+\frac{1}{2\alpha }\Bigl[ {\cal D}^mv^{(a)}_m\Bigr] ^2
-\bar{\eta}^{(a)}\Box \eta ^{(a)}
\label{4,3}
\end{eqnarray}
Here $\alpha $ is the gauge parameter, and $\eta ^{(a)}$ and
$\bar{\eta }^{(a)}$
are a ghost and an anti-ghost, respectively. As we can find easily,
even if we define that
both the ghost and the anti-ghost are SUSY invariant, the gauge fixing term
breaks the SUSY
(accurately, SUSY in Wess-Zumino gauge) invariance of total Lagrangian
$(\ref{4,3})$.  We know that the effective action ($W[g_{mn}]$) obtained
from this
Lagrangian is gauge parameter independent in Abelian case $\cite{b&v}$,
 however, so we could take the limit of $\alpha \to \infty$. This means, if
we take the limit which SUSY breaking gauge fixing term goes to zero in, then
the renormalizable Lagrangian $(\ref{4,3})$ is invariant under the
SUSY-transformation.

After all, we must do is to sum up the contributions of massless vector fields
with four degrees of freedom, massless Majorana spinor fields with two degrees
of freedom, and a pair of minimal massless ghost and anti-ghost field.

For a vector field with a ghost and an anti-ghost (totally, a gauge field),
the first two Seeley-DeWitt coefficients
$\cite{dw}$ are
\begin{eqnarray}
& &a_0=2
\label{4,4}\\
& &a_1=\frac{2}{3}{\cal R}
\label{4,5}
\end{eqnarray}
and for a Majorana spinor (gaugino) field
\begin{eqnarray}
& &a_0=2
\label{4,6}\\
& &a_1=\frac{1}{6}{\cal R}.
\label{4,7}
\end{eqnarray}
 Finally we obtain the induced Einstein action from an SU($N$) vector
multiplet
\footnote{In this context we ignored the infrared divergences appearing as
$m\to 0$ limit of eqs.$(\ref{2,14})$ and $(\ref{2,15})$.} which
is equivalent to $(N^2-1)$ U(1) vector multiplets
\begin{eqnarray}
W_V=\int d^4x\sqrt{g}\frac{1}{2(4\pi)^2}\Bigl[ -\frac{1}{2}{\cal R}\Lambda ^2
\Bigr](N^2-1) .
\label{4,8}
\end{eqnarray}
Then we find that with vector multiplets alone we always obtain negative
Newton constant, which is opposite to scalar multiplet case.


\section{Total Induced Einstein Action and Generations of Quarks,
Leptons and Their Superpartners}

Since we have no Einstein action at the original action,
summing up the two kinds of contributions obtained in the previous two
chapters, the total induced Einstein action is written as
\begin{eqnarray}
W=n_SW_S+n_VW_V
\label{5,1}
\end{eqnarray}
where $n_S$ and $n_V$ are non-negative integers.
 From this expression, again, we easily find that there is no
contribution to the cosmological constant through the vacuum fluctuations of
gauge and matter fields (including their superpartners).
It just means SUSY, if it is global or local, works.
However, the contribution to the
curvature term is nonvanishing in general. It decides the induced Newton
constant
$G_{ind}$.

Note that, while we studied thories with both gauge symmetries and SUSY
unbroken in the previous chapter,
even if we do the same with theories with gauge symmetry\footnote{In
this paper we mean QED or SU(N) gauge theory by gauge theory.}or SUSY
spontaneously broken taking into account the couplings to other scalar
multiplets, we obtain the same
results.
This is because the relation of mass squareds, which in general depend on
calssical fields, summed over all the species
contained in supersymmetric Lagrangian:
\begin{eqnarray}
\mbox{Str}\; m^2\equiv \sum _BE_{0B}m^2_B-\sum _FE_{0F}m^2_F=0
\label{4,9}
\end{eqnarray}
where subscripts $B$ and $F$ denote all the kinds of bosons and fermions,
respectively, holds even after spontaneous symmetry breaking
$\cite{w&b}$. In the above eq.$(\ref{4,9})$ $E_{0B}$ or $E_{0F}$ is
a coefficient appearing in eq.$(\ref{2,15})$.
 Then the above relation prevents mass dependent quadratically divergent
term appearing\footnote{ See eq.$(\ref{2,15})$}, if gauge symmetry or SUSY is
spontaneously broken or not.

We can easily find from eq.$(\ref{5,1})$ there are two cases depending on
the number of scalar
multiplets and one of vector multiplets. Note that we mean a multiplet of
a complex scalar and a Majorana spinor\footnote{ For scalar multiplets
we can use chiral spinors instead of Majorana ones} by a scalar multiplet,
and that of a gauge field with two polarizations and
a Majorana spinor by a vector multiplet. For example, we
count three for an SU(2) vector multiplet.
So the critical ratio of the number of scalar multiplets to one of vector
multiplets where the coefficient of Newton constant vanishes is three to
one.\footnote{ We do not care about the case that
the number of scalar multiplets is just three times as many as one of vector
multiplets, where the quadratically divergent contribution to the Newton
constant also vanishes and the logarithmically divergent one becomes
dominant, because it is too accidental and the coefficient of the
induced Einstien action depends on masses of fields contained in some
underlying theory $(\ref{2,15})$.} We investigate each of two cases in the
following two paragraphs.

(1) If the number of scalar multiplets is larger than three times as that of
vector multiplets, then
\begin{eqnarray}
\frac{1}{16\pi G_{ind}}=\mbox{ positive const.}
\times \Lambda ^2.
\label{5,2}
\end{eqnarray}
With $\Lambda =M_P\approx 10^{19}{\rm GeV}$ we obtain the positive induced
Newton
constant with same order of magnitude as ordinary Newton constant.

(2) If the number of scalar multiplets is smaller than three times as that of
vector multiplets, then
\begin{eqnarray}
\frac{1}{16\pi G_{ind}}=\mbox{negative const.}
\times \Lambda ^2
\label{5,3}
\end{eqnarray}
 In this case, we obtain negative induced Newton constant which means
repulsive
gravity contrary to Newton's theory of gravity.

We next apply this result to the supersymmetric unified models of elementary
particle. We pick up the two models which are hopeful candidates for 'beyond
the standard model'. One is minimal supersymmetric standard model (MSSM)
and another is minimal supersymmetric SU(5) GUT (MS SU(5) GUT), which is the
famous model as 'beyond the standard model' because of the precise
unification of three forces at so called GUT scale$\cite{l&l}$.
So, MS SU(5) GUT is important as a beyond the MSSM.

\subsection{Minimal supersymmetric standard model}

First, we count the number of scalar multiplets and vector multiplets
contained in MSSM
$\cite{f}$.

\begin{tabbing}
For scalar \= multiplets,                                                 \\
           \>$\bullet $ \= quarks and squarks: (a left-handed SU(2) doublet  \\
           \>  \>plus two right-handed SU(2) singlets) times three colors a
             generation,                                                  \\
           \>$\bullet $ \= leptons and sleptons: a left-handed SU(2) doublet \\
           \>  \>plus a right-handed SU(2) singlet a generation,             \\
           \>$\bullet $ Higgses and Higgsinos: two SU(2) doublets.          \\
For vector \= multiplets,                                                 \\
           \>$\bullet $ \= gauge fields and gauginos: an SU(3) octet         \\
           \>  \>plus an SU(2) triplet plus a U(1) gauge multiplet.          \\
\end{tabbing}

Totally, for $N_g$ generations there are (15$N_g$+4) scalar multiplets and 12
vector multiplets.
For $N_g=3,4,5,\ldots$ the case (1) realizes, otherwise the case (2) does.
So, if $N_g$ is equal to or larger than three, then gravity is attractive
definitly.

\subsection{Minimal supersymmetric SU(5) GUT}

Next, we do the same in MS SU(5) GUT$\cite{d&g}$.

\begin{tabbing}
For scalar \= multiplets,                                                   \\
           \>$\bullet $ \= quarks, leptons and their superpartners: a
${\bf 10}$ and a ${\bf \bar{5}}$ of SU(5)                                   \\
           \>  \>      a generation,                   \\
           \>$\bullet $ Higgses and Higgsinos: a ${\bf 24}$ and
 ${\bf 5}+{\bf \bar{5}}$ of SU(5).
     \\
For vector \= multiplets,                                                  \\
           \>$\bullet $ gauge fields and gauginos: a ${\bf 24}$ of SU(5).  \\
\end{tabbing}

 Numbers in bold face represent the dimensionality of representations under
SU(5), so they just mean the number of scalar multiplets and vector
multiplets.
Totally, for $N_g$ generations there are (15$N_g$+34) scalar multiplets and 24
vector multiplets.  The case of MS SU(5) GUT leads to the same result as MSSM
case.

\vspace{1cm}

Finally, we conclude that we need three or more generations of quarks,
leptons and their superpartners in order to make gravity attractive.
Of course, we can apply the
general result to any other models. For example, the extension of the above
 models with right-handed neutrinos and their superpartners also need
three generations or more to have an attractive gravity.


\section{Conclusion}

In this paper we investigated the induced Einstein gravity, or bosonic part
of induced Einstein supergravity, in supersymmetric unified models in general
space-time.
We considered that the gravity is purely induced from the general
covariant and supersymmetric unified model with
auxiliary gravity multiplet, namely neither the graviton nor the gravitino
is elementary but the composite object of quarks, leptons, gauge fields
etc..
Then we found that the sign of induced Newton constant depends on the ratio
of the number of scalar multiplets to that of gauge (vector) multiplets.
Applying the result to MSSM and MS SU(5) GUT, we concluded that we need
three or more generations of quarks and leptons and their superpartners in
order
to make gravity attractive.
\section*{Acknowledgements}

We would like to thank N. Maekawa, K. Tobe and T. Yanagida for giving us
useful comments and suggestions.

\end{document}